# Size-strain characteristics of lead and gold under fast ramp compression

E. F. Talantsev[1,2,†] and D. A. Komkova[1]

[1]M.N. Miheev Institute of Metal Physics, Ural Branch, Russian Academy of Sciences,

18, S. Kovalevskoy St., Ekaterinburg, 620108, Russia

[2]NANOTECH Centre, Ural Federal University, 19 Mira St., Ekaterinburg 620002, Russia

[†]e-mail evgeny.f.talantsev@gmail.com

**Abstract**

Phase transitions in materials under fast ramp compression are an ongoing research topic, which is part of several global projects like inertial fusion. Currently, X-ray diffraction (XRD) examination of samples under fast ramp compression is limited to the determination of the sample phase state and the unit cell lattice parameters. Here, we propose to extend this examination route by introducing the Williamson-Hall analysis of the XRD data measured in samples under fast ramp rate conditions. To demonstrate the applicability of the method, we performed analysis for ramp compressed lead ($P$=200 GPa) and gold ($P$=1003 GPa), which both exhibit a transition from the face-centred cubic (fcc) lattice to the body-centred cubic (bcc) lattice at the studied pressures. The analysis showed that lead under fast ramp compression has a nanocrystalline structure with a crystalline size of $D$=(4±1) nm and lattice strain $\varepsilon$=(0.6±0.2)%. The effect of extreme hardening of *bcc*-Pb under fast ramp compression can be explained as the formation of an ultrafine grain structure in this metal. Elemental gold exhibits average crystalline size $D$>12 nm and unprecedentedly high, for a pure metallic element, lattice strain $\varepsilon$~1.5% under fast ramp compression.

# Size-strain characteristics in lead and gold under fast ramp compression

## 1.     Introduction

Face-centred cubic (fcc) to the hexagonal close-packed (hcp) polymorphic phase transition in lead had been discovered more than five decades ago [1]. At higher pressures and under fast ramp compression, elemental lead exhibits the hcp to the body-centred cubic (bcc) polymorph transition [2–5]. Recently, Coleman et al [6] reported on experimental observation of the fcc → bcc polymorph phase transition in pure gold under fast ramp compression at pressures above 1 TPa.

Besides being a condition for the polymorph phase transitions, the fast ramp compression also a condition to observe the effect of extreme hardening of pure metals [4,7]. Dowding and Schuh [7] explained this effect as a manifestation of the absence of ductile mechanisms in metals under fast ramp compression, since the motion of defects (such as dislocations) remains inactive under fast ramp compression conditions, as the defects in the crystalline samples remain immobile due to the existence of a minimal and unchangeable internal lower kinetic limit for the response of these defects to the applied strain. In other words, the impact is so rapid that the structural defects in crystalline solids do not have enough time to begin moving in response to rapidly increasing deformation.

For decades [8–13], X-ray diffraction (XRD) data analysis of fast ramp compressed materials was limited to the determination of the material unit cell parameters and the lattice symmetry. Recently, Coleman et al [14] extended the analysis with a proposal to extract the density of randomly distributed stacking faults from the shape asymmetry of the reflection peaks (which is the standard approach in XRD studies [15–18]) and applied this method to fast ramp compressed silver.

Here, in an attempt to further advancing tools used for the analysis of XRD data measured in fast ramp compressed materials, we propose to use the Williamson-Hall analysis [19].

In the result, we found that pure lead (at fast ramp compression at $P$ = 200 GPa) exhibits the ultrafine grain structure with an average crystalline size of $D$ = (4 ± 1) nm and high (for ductile pure metallic lead) lattice strain ε = (0.6 ± 0.2)%. The revealed ultrafine grain structure in fast ramp compressed lead explains the effect of extreme hardening of pure lead under fast ramp compression [4], because the extreme hardening effect [20] is observed in pure metals with ultrafine grain structure under normal conditions.

We also found that *fcc*-phase and *bcc*-phases of gold (at fast ramp compression at $P$ = 1003 GPa) exhibit average crystalline sizes $D$ that exceed the resolution of the XRD diffraction facility [8] used in fast ramp compression experiments. However, we deduced the lattice strains in *fcc*- and *bcc*-phases these phases, which have an unprecedentedly high value (for pure ductile gold metal) ε ~ 1.5%.

## 2. Experimental data sources and fitting functions

In this study, we analysed experimental data for fast ramp compressed lead (XRD data reported by Rygg *et al*. [8]) and gold (XRD data reported by Coleman et al [6]). Experimental curves showed in Figs. 14, 21 in Ref. [8] and Figs. 2, S2 in Ref. [6] were digitized.

Williamson-Hall analysis [19] is based on the determination of the average size of the crystallites $D$ and the micro-deformation in these crystallites ε from the fit of experimental integral breadth $\beta\left(\frac{2\theta}{2}\right)$ of the diffraction peaks observed at the Bragg's diffraction angle $2\theta_B$ to the equation:

$$\beta(2\theta_B) = \frac{0.9 \times \lambda_{X-ray}}{D \times cos\left(\frac{2\theta_B}{2}\right)} + 4 \times \varepsilon \times tan\left(\frac{2\theta_B}{2}\right), \tag{1}$$

where $\lambda_{X-ray} = 121.55\ pm$ is the X-ray wavelength used in Ref. [6,8]. We performed data fits to all equations in this study by using the Origin software [21].

A profile of each measured in the experiment Bragg's reflection was fitted to the Gaussian peak function [8]:

$$I(2\theta) = I_{background} + \frac{A}{\sigma(2\theta) \times \sqrt{2\pi}} \times exp\left(-\frac{(2\theta - 2\theta_B)^2}{2 \times (\sigma(2\theta))^2}\right), \quad (2)$$

where $I_{background}$, $A$, $2\theta$, and $\sigma$ are free-fitting parameters, and the integral breadth $\beta(2\theta)$ of the diffraction peak is calculated by the following equation:

$$\beta(2\theta) = \sqrt{2\pi} \times \sigma(2\theta) \quad (3)$$

Fits for all X-ray diffraction peaks that were analysed in this study are shown in the Supplementary Information Figs. S1, S2.

## 3. Instrumental broadening of the XRD diffractometer used in fast ramp compression experiments

Rygg et al [8] reported spectral and total instrumental broadening function $\sigma_i(2\theta)$ (in their Fig. 21 in Ref. [8]) for the diffractometer used in fast ramp compression experiments [6,8]. All datasets analysed in this study were obtained on this diffractometer, and reported $\sigma_i(2\theta)$ data [8] is shown in Fig. 1,a.

The shape of the $\beta_i(2\theta)$ function (Fig. 1) differs from any instrumental broadening function $\beta_i(2\theta)$ reported for laboratory X-ray tube machines and synchrotrons [22–30]. Based on that, widely used Caglioti- Paoletti-Ricci function [31] to approximate the $\sigma_i(2\theta)$ and $\beta_i(2\theta) = \sqrt{2\pi} \times \sigma_i(2\theta)$ gives inaccurate approximation.

Our examination of the experimental XRD datasets reported by this group [8] in other reports showed that the value of $\sigma_i(2\theta)$ reported in Fig. 21 of the Ref. [8] is significantly overestimated. Indeed, the $\sigma(2\theta)$ of the Bragg reflections for platinum pinholes in the XRD datasets for fast ramp compressed gold [6] are significantly narrower than the $\sigma_i(2\theta)$ in Fig. 21 of Ref. [8]. Thus, here we use the Bragg peaks from Pt pinholes in XRD datasets of fast ramp compressed gold at pressure $P = 545$ GPa [6] as the peaks from profile shape standard sample. The approximation of these peaks to Eq. 2 and deduced parameters are shown in Fig. S1. Deduced $\sigma_{Pt}(2\theta)$ and $\beta_{Pt}(2\theta) = \sqrt{2\pi} \times \sigma_{Pt}(2\theta)$ are shown in Fig. 1 together with the fits of these datasets to the Caglioti- Paoletti-Ricci function [31]:

$$\beta_{inst}(2\theta) = \sqrt{U \times tan^2\left(\frac{2\theta}{2}\right) + V \times tan\left(\frac{2\theta}{2}\right) + W}, \quad (4)$$

where $U$, $V$, and $W$ are free-fitting parameters, for which deduced values are shown in Fig. 1.

It should be noted that only at $2\theta \geq 100\ degree$ the data presented by Rigg et al. [8] and the extrapolation by the Caglioti-Paoletti-Ricci method [31] of the data obtained by us coincide (Fig. 1). We also need to point out that true instrumental broadening should be below the level we establish for the platinum herein, and precise experiments for better determination of the instrumental broadening of the diffractometer used in fast ramp compression experiments are needed.

We can calculate the resolution of the diffractometer by utilizing a primary idea of practical physics [32], where the resolution of the measuring device is determined by the half of the minimal division of the scale. In our case (Fig. 1), this is half of the integral peak breadth of platinum peaks at $2\theta = 31.09°$, that is $\beta_{min} = \frac{\beta_{Pt}(2\theta=31.09°)}{2} = \frac{1.918 \times 10^{-2}\ rad}{2} = 9.59 \times 10^{-3}\ rad$. Substituting this value of $\beta_{min}$ in Scherrer equation [33], one can obtain:

$$D_{max} = \frac{0.9 \times \lambda_{X-ray}}{\beta_{min} \times cos\left(\frac{2\theta}{2}\right)} = 12\ nm. \quad (5)$$

Despite calculated $D_{max}$ value (Eq. 5) is dozens of times lower than typical $D_{max}$ value in usual laboratory XRD machines [16,30,33], it should be taken into account extreme conditions at which fast ramp experiments are performing.

To determine the broadening of the XRD peaks that originates from the sample, $\beta_s(2\theta)$, we used the standard equation for Gaussian functions [23–25,30]:

$$\beta_s(2\theta) = \sqrt{\beta^2(2\theta) - \beta_{Pt}^2(2\theta)} \quad (6)$$

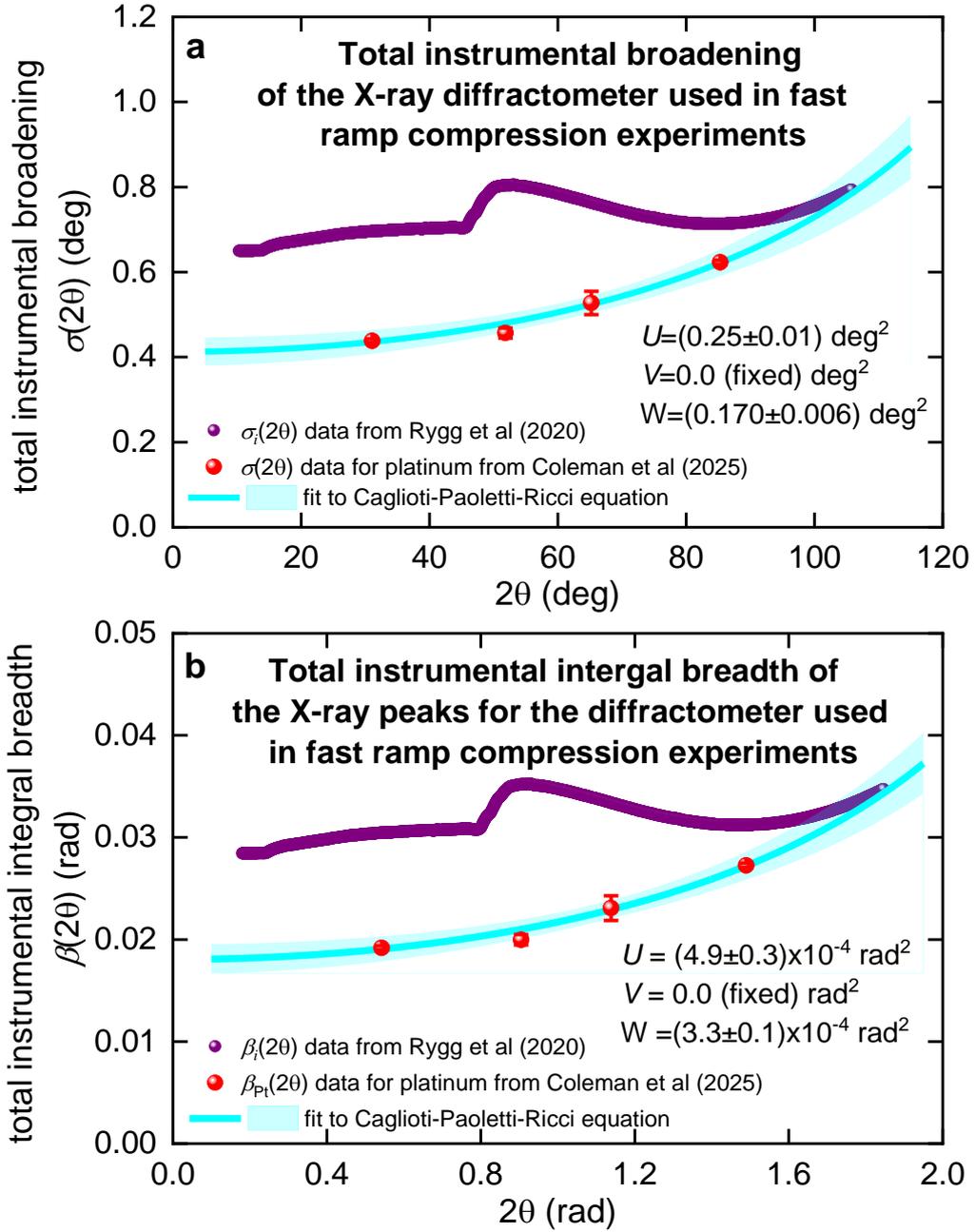

**Fig. 1.** Total instrumental broadening datasets (**a**) $\sigma_i(2\theta)$ and (**b**) $\beta_i(2\theta)$ for the diffractometer used in fast ramp compression experiments by Rygg et al [8]. Also, data of (**a**) $\sigma_{Pt}(2\theta)$ and (**b**) $\beta_{Pt}(2\theta)$ deduced for platinum pinholes in experiments by Coleman et al [6] for gold compressed at $P$ =545 GPa are shown. Data of (**a**) $\sigma_{Pt}(2\theta)$ and (**b**) $\beta_{Pt}(2\theta)$ were fitted to Caglioti- Paoletti-Ricci function [31] and deduced parameters are: (**a**) $U = (0.25 \pm 0.01)\ degree^2, V \equiv 0\ (fixed)\ degree^2; W = (0.170 \pm 0.006)\ degree^2$; (**b**) $U = (4.9 \pm 0.3) \times 10^{-4}\ radian^2, V \equiv 0\ radian^2; W = (3.3 \pm 0.1) \times 10^{-4}\ radian^2$. Fits quality (R-Square (COD)) is 0.9943. 95% confidence band is shown by the cyan area.

## 4. Strain-size dependence in fast ramp compressed lead

The fit of the $\beta_s(2\theta)$ data to Williamson-Hall equation (Eq. 1) for fast ramp compressed lead (at $P$ = 200 GPa) is shown in Fig. 2. Deduced average crystalline size is $D$ = (4.3 ± 0.7) nm, and lattice strain in these nanocrystals is $\varepsilon$ = (0.6 ± 0.2)%.

Deduced parameters reaffirm the existence of the effect of the extreme hardening in lead under fast ramp compression [4,7], because the effect of high-pressure strengthening in ultrafine-grained pure metals (at static pressures and crystalline size of ~ 3 nm) is a well-known phenomenon [20]. Besides small grain size, the deduced strain in lead nanocrystals $\varepsilon$ = (0.6 ± 0.2)% is unprecedentedly high for ductile pure metallic lead. In the result, we can confirm that based on deduced strain-size characteristics of fast ramp compressed lead, the effect of extreme hardening in lead at fast ramp compression [4,7] is expected from basic structural parameters of this compressed metal.

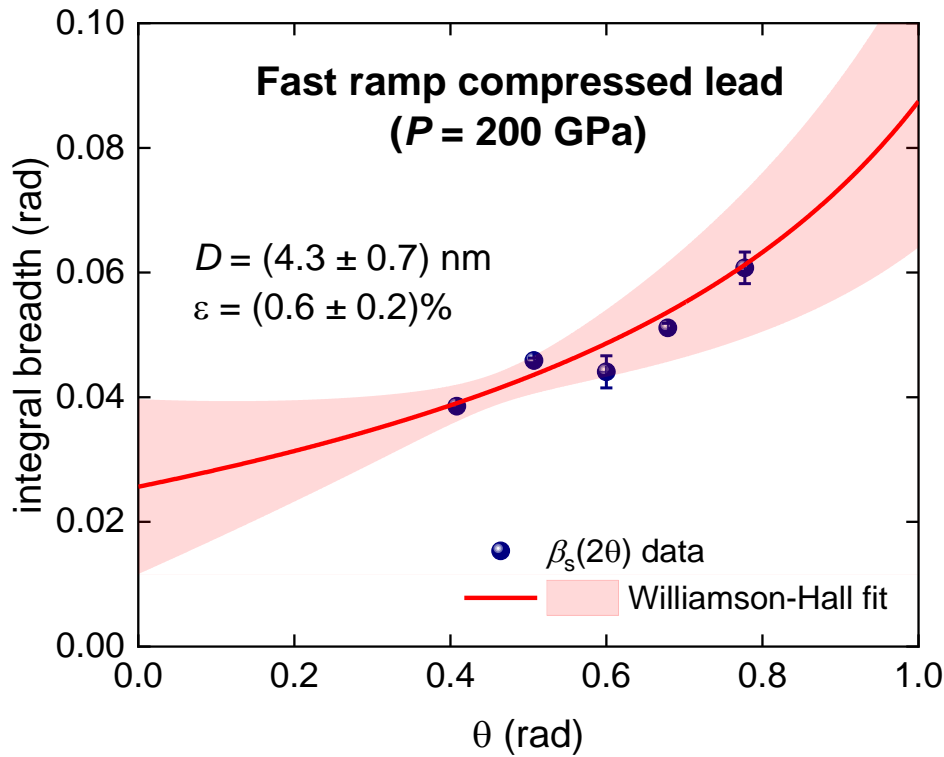

**Fig. 2.** The fit of the $\beta_s(2\theta)$ data for fast ramp compressed lead at $P$ = 200 GPa to the Williamson-Hall equation (Eq. 1). Experimental XRD dataset from which the data was reported by Rygg et al [8] (in their Fig. 14 [8]). Deduced parameters are shown in the plot. Fit quality (R-Square (COD)) is 0.8945. 95% confidence band is shown by the pink area.

## 5. Strain-size dependence in fast ramp compressed gold

Coleman et al [6] reported that elemental gold (that exhibits *fcc*-lattice at normal conditions) transforms into a two-phase *fcc+bcc* state under fast ramp compression at *P* = 1003 GPa. We fitted XRD peaks to Eq. 2 under the designation of peaks proposed by Coleman et al [6] for the *fcc*-lattice and *bcc*-lattice. Fits are shown in Fig. S3.

The obtained $\beta_s(2\theta)$ dataset for each phase was fitted to Williamson-Hall equation (Eq. 1) and these datasets and fits are shown in Fig. 3.

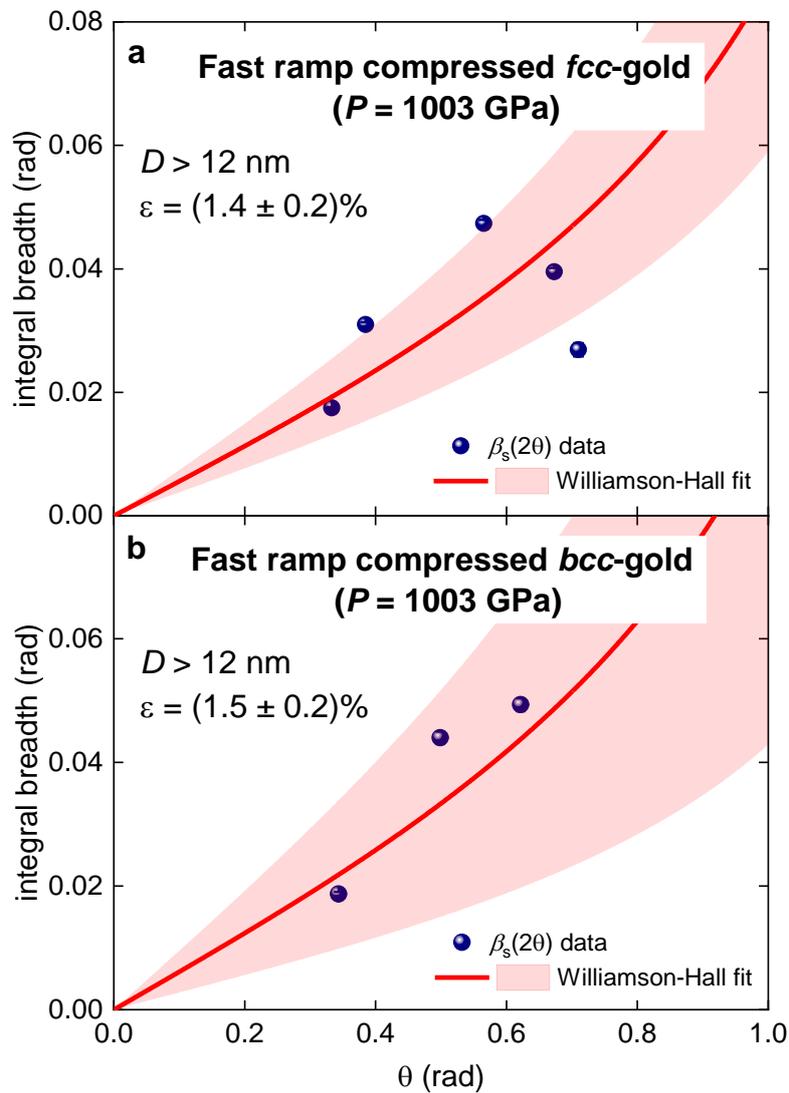

**Fig. 3.** The fit of the $\beta_s(2\theta)$ data for fast ramp compressed gold at *P* = 1003 GPa to the Williamson-Hall equation (Eq. 1). Experimental XRD dataset from which the data was extracted reported by Coleman et al [6] (in their Fig. 2 [6]). (**a**) data and fit for *fcc*-gold, fit quality (*R*-Square (COD)) is 0.6422; (**b**) data and fit for *bcc*-gold. Deduced parameters are shown in the plot. Fit quality (*R*-Square (COD)) is 0.8154. 95% confidence band is shown by the pink areas in both panels.

Both phases exhibit average crystalline sizes that exceed the resolution of the diffractometer (Eq. 5), $D_{fcc-Au} > D_{max} = 12\ nm$ (Fig. 3,a) and $D_{bcc-Au} > D_{max} = 12\ nm$ (Fig. 3,b). The deduced lattice strain ε (in both phases) is unprecedentedly high ($\varepsilon_{fcc-Au} = (1.4 \pm 0.2\ )\%$ and $\varepsilon_{bcc-Au} = (1.5 \pm 0.2\ )\%$) for extremely ductile pure gold at normal conditions.

## 6. Discussion

It needs to be mentioned that recently Williamson-Hall approach [19] was applied for the XRD data analysis for compressed superconductors in diamond anvil cells (DAC) [27–29,34,35], including elemental sulfur [29] (XRD data was reported by Du et al [36]), $H_3S$ [34] (XRD data was reported by Du et al [37]), $La_3Ni_2O_{7-\delta}$ [28] (XRD data was reported by Sun et al [38]), $La_4H_{23}$ (XRD data was reported by Guo et al [39]), and $BaH_{12}$ [35] (XRD data was reported by Chen et al [40]). Also, Williamson-Hall approach [19] was applied for analysis data for sulfur chains [29] formed by fast compression method [41] for samples in DAC [42–46].

In this study, we have made the next step and extended the Williamson-Hall approach [19] on fast ramp compressed experiments [8]. Despite a relatively low spatial resolution of the XRD facility [8], where the upper crystalline size limit is $D_{max} = 12\ nm$, we show that fast ramp compressed lead exhibits the average crystalline size of $D{\sim}4\ nm$. And therefore, the effect of extreme hardening of lead under fast ramp compression can be explained as a direct consequence of the formation of the ultrafine grain structure.

This finding explains the effect of extreme hardening of lead under fast ramp compression [4]. Our explanation differs from the basic idea of Dowding and Schuh [7,47], who suggested that the effect is a manifestation of the absence of plastic mechanisms in metals under fast ramp compression, since the movement of defects (such as dislocations) remains inactive under conditions of fast ramp compression, as the defects in the crystalline samples remain stationary because of the existence of a minimum internal lower time limit for the response of these defects to the applied deformation.

Our results (Fig. 2) showed that elemental lead exhibits an ultrafine grain structure with an average grain size of $D = 4$ nm. This ultrafine grain structure is formed in lead under fast ramp compression conditions. And as a direct consequence of this formation, the effect of extreme hardening should emerge, similar to the strengthening effect observed in ultrafine-grained metals at ambient conditions.

## 7. Conclusion

In this study, we propose to use Williamson-Hall method [19] for the analysis of the XRD data measured in fast ramp compressed materials. Performed analysis of X-ray diffraction data measured in fast ramp compressed samples of elemental lead and gold has shown the advantages of using this approach for studying materials under extreme conditions.

*Supplementary Materials*

In Supplementary Materials we showed all Bragg's peaks fits to Gaussian function (Eq. 2) that were analyzed in this study.

*Acknowledgements*

The work was carried out within the framework of the state assignment of the Ministry of Science and Higher Education of the Russian Federation for the IMP UB RAS. E.F.T gratefully acknowledged the research funding from the Ministry of Science and Higher Education of the Russian Federation under Ural Federal University Program of Development within the Priority-2030 Program.

# Supplementary Materials

# Size-strain characteristics of lead and gold under fast ramp compression

## by E.F. Talantsev and D.A. Komkova

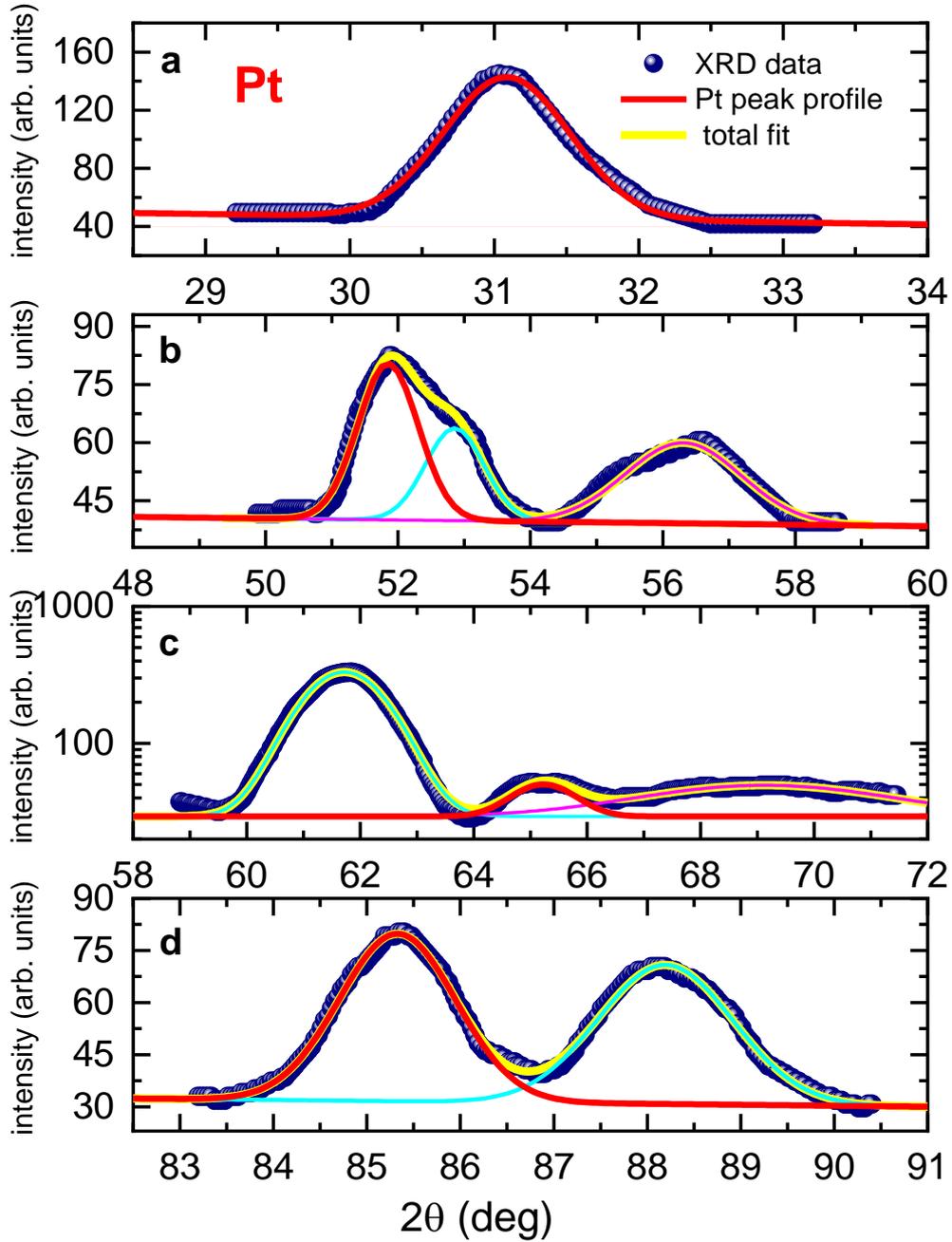

**Figure S1.** XRD data and data fits to the Gaussian function (Eq. 2 in main text, or Eq. 31 in Ref. [8]) for Pt and Au peaks measured in a sample of gold compressed at $P = 545$ GPa [6]. Fitting curves for peaks of platinum are in red. Deduced parameters for Pt are: (**a**) $2\theta = (31.087 \pm 0.004)\ deg$, $\sigma = (0.438 \pm 0.005)\ deg$, fit quality ($R$-Square (COD)) is 0.9938; (**b**) $2\theta = (51.85 \pm 0.04)\ deg$, $\sigma = (0.457 \pm 0.011)\ deg$, fit quality ($R$-Square (COD)) is 0.9883. (**c**) $2\theta = (65.23 \pm 0.03)\ deg$, $\sigma = (0.528 \pm 0.028)\ deg$, fit quality ($R$-Square (COD)) is 0.9986; (**d**) $2\theta = (85.329 \pm 0.004)\ deg$, $\sigma = (0.623 \pm 0.005)\ deg$, fit quality ($R$-Square (COD)) is 0.9957.

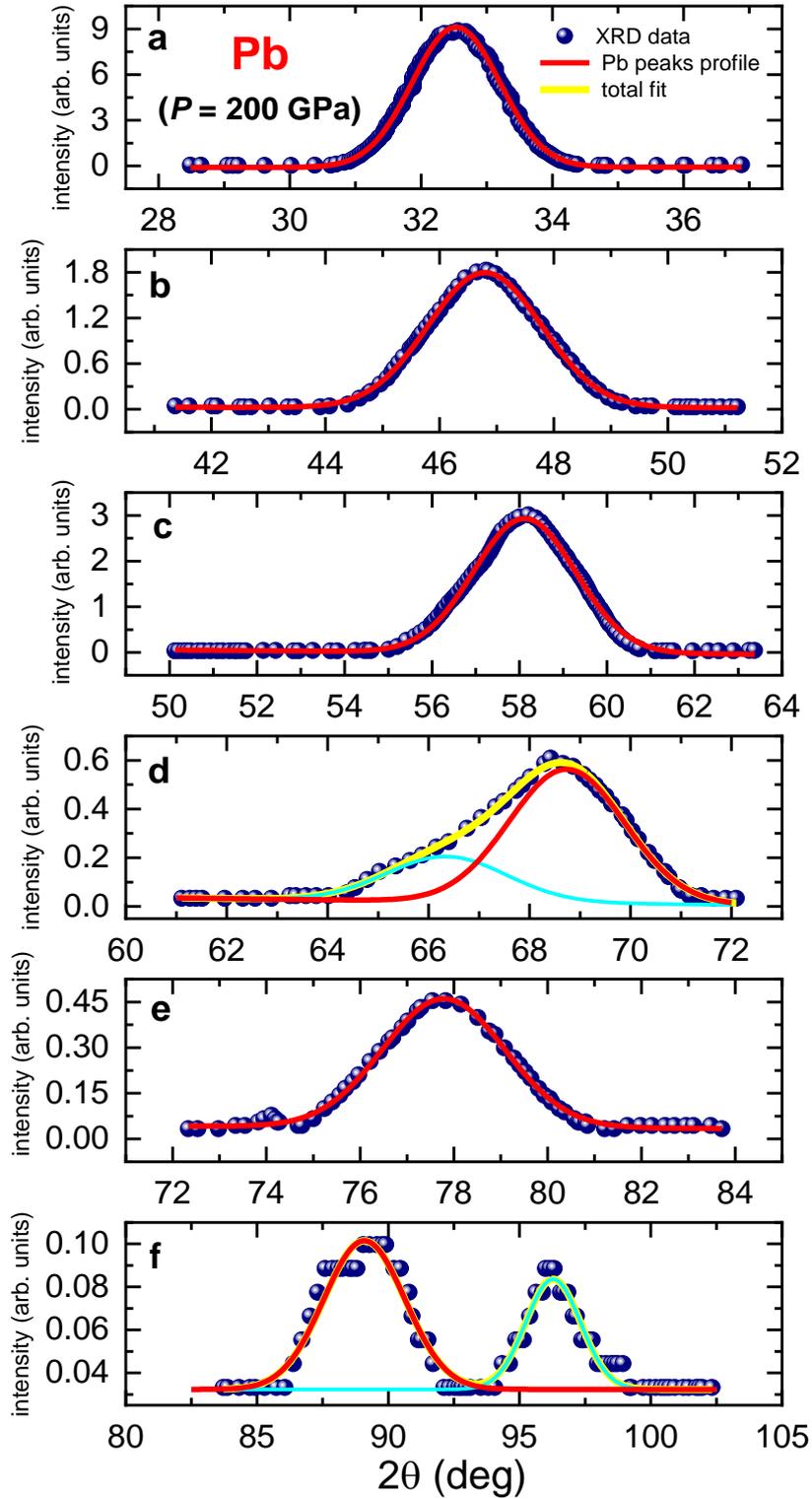

**Figure S2.** XRD data and data fits to the Gaussian function (Eq. 2 in main text, or Eq. 31 in Ref. [8]) for Pb peaks measured in a sample of pure lead compressed at $P = 200$ GPa [8]. Deduced parameters are: (**a**) $2\theta = (32.532 \pm 0.001)\ deg$, $\sigma = (0.668 \pm 0.002)\ deg$, fit quality ($R$-Square (COD)) is 0.9968; (**b**) $2\theta = (46.793 \pm 0.004)\ deg$, $\sigma = (0.998 \pm 0.005)\ deg$, fit quality ($R$-Square (COD)) is 0.9986. (**c**) $2\theta = (65.236 \pm 0.007)\ deg$, $\sigma = (1.163 \pm 0.009)\ deg$, fit quality ($R$-Square (COD)) is 0.9962; (**d**) $2\theta = (85.33 \pm 0.09)\ deg$, $\sigma = (1.14 \pm 0.06)\ deg$, fit quality ($R$-Square (COD)) is 0.9975; (**e**) $2\theta = (77.78 \pm 0.01)\ deg$, $\sigma = (1.31 \pm 0.02)\ deg$, fit quality ($R$-Square (COD)) is 0.9942. (**f**) $2\theta = (89.09 \pm 0.05)\ deg$, $\sigma = (1.55 \pm 0.06)\ deg$, fit quality ($R$-Square (COD)) is 0.9569.

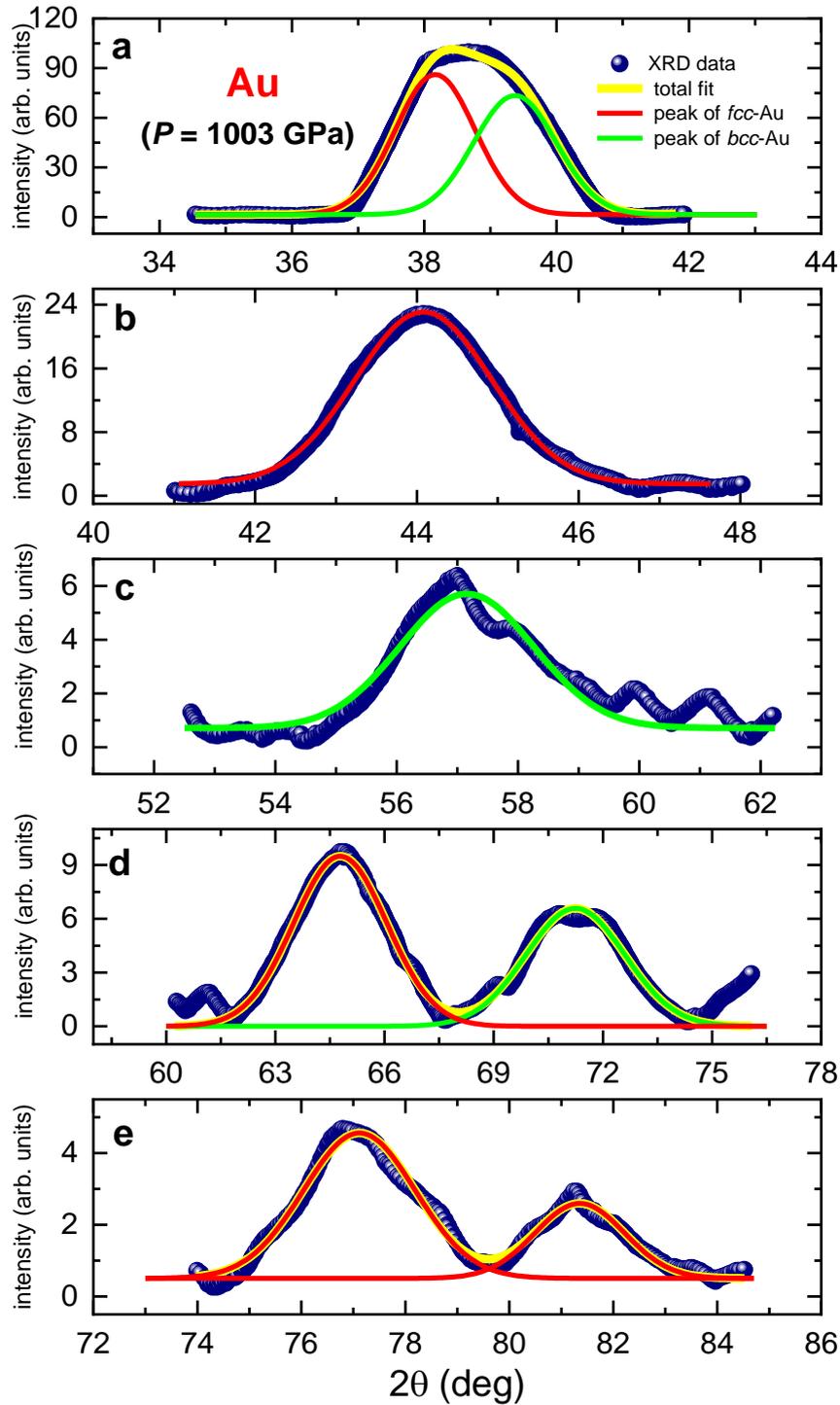

**Figure S3.** XRD data and data fits to the Gaussian function (Eq. 2 in main text, or Eq. 31 in Ref. [8]) for Au peaks measured in a sample of pure gold compressed at $P$ = 1003 GPa [6]. Fitting curves for *fcc*-Au peaks are in red, and for *bcc*-Au peaks are in green. Deduced parameters are: (**a**) *fcc*-Au: $2\theta = (38.166 \pm 0.013)\ deg$, $\sigma = (0.602 \pm 0.004)\ deg$; *bcc*-Au: $2\theta = (39.372 \pm 0.014)\ deg$, $\sigma = (0.623 \pm 0.008)\ deg$; fit quality (*R*-Square (COD)) is 0.9958; (**b**) *fcc*-Au: $2\theta = (44.081 \pm 0.008)\ deg$, $\sigma = (0.846 \pm 0.009)\ deg$, fit quality (*R*-Square (COD)) is 0.9980. (**c**) *bcc*-Au: $2\theta = (57.18 \pm 0.02)\ deg$, $\sigma = (1.12 \pm 0.02)\ deg$, fit quality (*R*-Square (COD)) is 0.9245; (**d**) *fcc*-Au: $2\theta = (64.789 \pm 0.009)\ deg$, $\sigma = (1.20 \pm 0.01)\ deg$; *bcc*-Au: $2\theta = (71.25 \pm 0.01)\ deg$, $\sigma = (1.26 \pm 0.02)\ deg$ fit quality (*R*-Square (COD)) is 0.9926. (**e**) *fcc*-Au: $2\theta = (77.12 \pm 0.01)\ deg$, $\sigma = (1.07 \pm 0.01)\ deg$; $2\theta = (81.36 \pm 0.02)\ deg$, $\sigma = (0.86 \pm 0.02)\ deg$ fit quality (*R*-Square (COD)) is 0.9786.